\documentclass[sn-mathphys,Numbered]{sn-jnl}

\usepackage{graphicx}%
\usepackage{multirow}%
\usepackage{amsmath,amssymb,amsfonts}%
\usepackage{amsthm}%
\usepackage{mathrsfs}%
\usepackage[title]{appendix}%
\usepackage{xcolor}%
\usepackage{textcomp}%
\usepackage{manyfoot}%
\usepackage{booktabs}%
\usepackage{algorithm}%
\usepackage{algorithmicx}%
\usepackage{algpseudocode}%
\usepackage{listings}%

\theoremstyle{thmstyleone}%
\theoremstyle{thmstyletwo}%

\theoremstyle{thmstylethree}%

\begin{document}

\title[Non-eq Dynamics]{Non-equilibrium Dynamics of Vortices in Two-Dimensional Quantum Gases: Determining the Dynamical Scaling Region Using the Mahalanobis Distance}


\author*[1]{\fnm{Richard} \sur{Tattersall}}\email{r.tattersall2@newcastle.ac.uk}

\author[1]{\fnm{Andrew} \sur{Baggaley}}\nomail

\author[1]{\fnm{Thomas} \sur{Billam}}\nomail

\affil[1]{\orgdiv{Joint Quantum Centre (JQC) Durham-Newcastle}, \orgname{School of Mathematics, Statistics and Physics, Newcastle University}, \orgaddress{\city{Newcastle upon Tyne}, \postcode{NE1 7RU}, \country{United Kingdom}}}

%


\abstract{When a two-dimensional system undergoes a rapid quench from a disordered to an ordered phase, it does not order instantly but instead relaxes towards equilibrium over time.  During this relaxation, the dynamical scaling hypothesis predicts that the length scale of ordered regions increases, with later patterns statistically similar to earlier ones except for a change in global scale.  Quantum gases are one of many systems in which such out of equilibrium behaviour is predicted to occur.  Here we present a method for systematically testing when dynamical scaling is taking place, by quantifying the similarity of the rescaled two-point correlation function over time using the Mahalanobis distance.  Data on the velocity field of a quantum fluid, generated from point vortex simulations, are used to illustrate the application of this method.}




\maketitle

\section{Introduction}\label{sec1}

People search for evidence of dynamical scaling in systems such as the Ising model of magnetic spins, binary alloys \cite{Bray1994,Nishimori2011} and Bose-Einstein Condensates (BECs) \cite{Groszek2021}.  A key prediction of the dynamical scaling hypothesis is that the correlation function, $G(r,t)$, between two points separated by a distance $r$, changes in time, $t$,  only by a rescaling in length scale by $L_{c}(t)$, where $L_{c}(t)$ is a correlation length that depends on time only.
\begin{equation}
G(r,t)=G(r/L_{c}(t))\,.
\label{eqn:scaling}
\end{equation}
To test if this prediction is met, the correlation functions at different times are rescaled by $L_{c}(t)$ and plotted on the same axes.  If they collapse onto a single curve, this is evidence of dynamical scaling.  Typically, this collapse is checked visually; however, a more rigorous, algorithmic test is preferable in order to determine the times at which dynamical scaling applies. In this paper we outline an approach that makes use of the Mahalanobis distance \cite{Mahalanobis1936_2018} to quantify the deviation of a curve from the mean, eliminate outliers and identify the scaling region.

\begin{figure}
	\centering
	\includegraphics[width=0.9\textwidth]{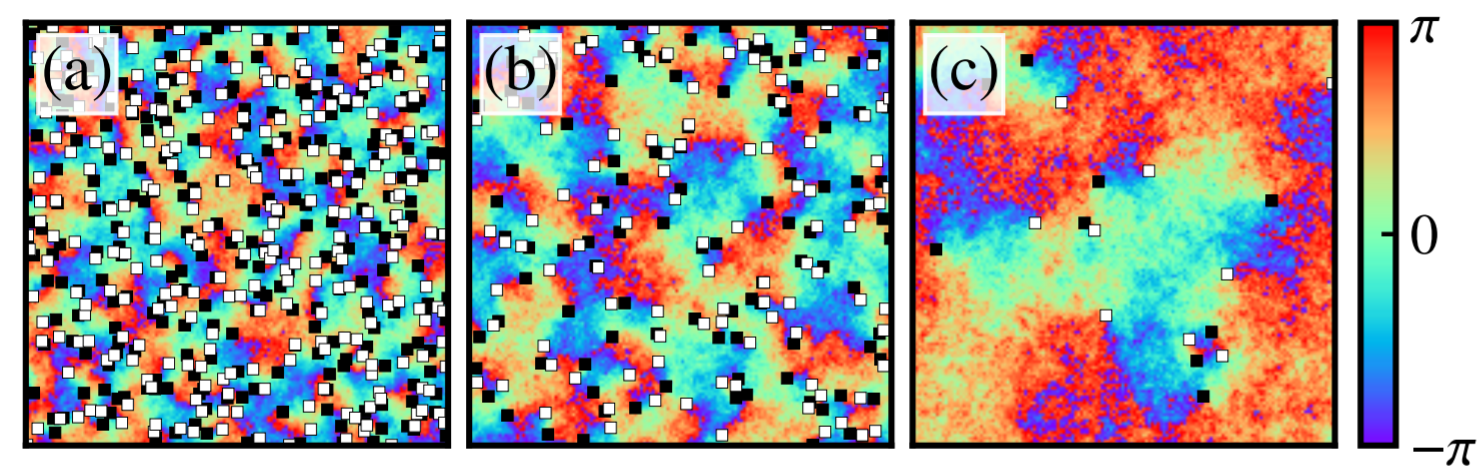}
	\caption{Evolution of the phase of the field $\psi$ of a BEC in the case of zero dissipation, reproduced from \cite{Groszek2021} under CC-BY-4.0 international license \protect\footnotemark.  Distances are quoted in terms of the healing length, $\xi=\hbar/\sqrt{m\mu}$, and times are given in terms of $\tau=\hbar/\mu$.  These are the characteristic length and time scales used when solving the Gross-Pitaevskii equation, where $m$ is the atomic mass, $\hbar$ is the reduced Planck constant and $\mu$ is the chemical potential.  The BEC was modelled using the projected Gross-Pitaevskii equation (PGPE) in a doubly-periodic square box of width $\approx 363\xi$.  The BEC was created in a far from equilibrium configuration with a high density of quantised vortices and antivortices.  It then ``relaxes'', in this case through annihilation of vortex-antivortex pairs. (a), (b) and (c) correspond to times $t \approx \{ 200,2000,20000 \}\tau $ respectively.  The open white squares show the positions of vortices (positive circulation), the filled black squares show the positions of antivortices (negative circulation) and the background colour indicates the phase.}\label{fig:coarsening}
\end{figure}

When dynamical scaling holds, the dependence of correlation length on time is predicted to be a power law of the form:
\begin{equation}
L_{c}(t)\sim t^{1/z}\,,
\label{eqn:corr_length_power_law}
\end{equation}
where $z$ is the dynamical critical exponent and is expected to have a value of two for a two-dimensional system with a non-conserved order parameter \cite{Bray1994,Nishimori2011}.  A recent numerical study of coarsening dynamics, following the rapid quench of a Bose gas into the ordered phase, found clear evidence of dynamical scaling \cite{Groszek2021}.  The values of $z$ found were close to two but dependent upon dissipation, parameterized by a dimensionless constant, $\gamma$.  This study used the projected Gross-Pitaevskii equation (PGPE) and the stochastic projected Gross-Pitaevskii equation (SPGPE) to model the evolution of the wavefunction $\psi(\mathbf{r},t)$, both in the absence and the presence of dissipation.  The authors observed growing regions of constant phase following the quench, as quantum vortices and antivortices annihilated, as shown in Fig.~\ref{fig:coarsening}.  They found the correlation functions at different times using the wavefunction and observed them to collapse, within a given scaling window, onto a single curve once rescaled by $L_{c}(t)$.  Their method for identifying the scaling window was to judge by eye initially and later check that the maximum discrepancy between rescaled correlation functions was sufficiently small.  An earlier study \cite{Karl2017}, simulating a similar system using a truncated Wigner approach, also found evidence of universal scaling behaviour in the time evolution of the single-particle momentum spectrum.  They selected a scaling window by inspecting the rescaled spectra and found a value of $z=1.8(3)$ (avoiding a strongly anomalous non-thermal fixed point associated with vortex clustering).     
To illustrate our method to test for the collapse of rescaled correlation functions, we also consider a two-dimensional quantum fluid.  However, rather than simulate the system using the Gross-Pitaevskii equation (GPE) \cite{Pitaevskii2016,Pethick2008}, we use the point vortex model (PVM) instead \cite{Helmholtz1858,Kirchoff1876}.  The PVM is often used to simulate dynamics in two-dimensional quantum fluids \cite{Fetter1966,Billam2015,Skipp2023} such as those formed in ultracold atomic gases \cite{Anderson1995,Abo-Shaeer2002}.  Although vortices in atomic gases are not point-like and the gas itself is compressible, the PVM can give reasonably accurate predictions for vortex behaviour.  It also benefits over the more directly applicable GPE because only the coordinates of the vortices need tracking and not the quantum mechanical wavefunction at every grid-point.  Consequently, it is computationally less expensive and makes simulations of larger systems possible.  
\footnotetext{https://creativecommons.org/licenses/by/4.0/}
The structure of this paper is as follows: Section~\ref{sec_meth} describes how data for the out of equilibrium dynamics of a two-dimensional quantum fluid is generated using a dissipative PVM.  Section~\ref{sec_analysis} details how a two-point correlation function is calculated and rescaled and how the scaling region is determined using a method based on the Mahalanobis distance.  Section~\ref{sec_conc} summarises the main findings and suggests some possibilities for future work using this method.   

\section{Generating Data Using the Point Vortex Model (PVM)}\label{sec_meth}
\subsection{The PVM on an Infinite Plane}\label{subsec_PVM}
The PVM applies to two-dimensional incompressible fluids with zero viscosity and assumes that vorticity is confined to delta-function-like points \cite{Helmholtz1858}.  Each vortex is advected by its local fluid velocity, which is made up of contributions from each of the other vortices.  Consequently, the equations of motion for vortex $i$, of circulation, $\kappa_{i}$, at position $(x_{i},y_{i})$, in a system of $N_{v}$ point vortices are:  
\begin{equation}
\frac{d x_{i}}{dt} =-\frac{1}{2\pi}\sum_{j=1,j\neq i}^{N_{v}}\frac{\kappa_{j}y_{ij}}{|\mathbf{r}_{ij}|^2}\,,\,\,
\frac{d y_{i}}{dt} =\frac{1}{2\pi}\sum_{j=1,j\neq i}^{N_{v}}\frac{\kappa_{j}x_{ij}}{|\mathbf{r}_{ij}|^2}\,.
\label{eqn:pv_eqm}
\end{equation}
Where $\kappa_{j}$ and $(x_{j},y_{j})$ are the circulations and positions of all the vortices other than $i$ and $x_{ij}=x_{i}-x_{j}$, $y_{ij}=y_{i}-y_{j}$ and $\mathbf{r}_{ij}=(x_{ij},y_{ij})$.  This result may also be derived from the Hamiltonian \cite{Kirchoff1876}:
\begin{equation}
H=-\frac{1}{4\pi} \sum_{i=1}^{N_{v}}\sum_{j=1,j\neq i}^{N_{v}}\kappa_i\kappa_j \ln{|\mathbf{r}_{ij}|}\,,
\label{eqn:pv_H}
\end{equation}
in which the double sum adds contributions from all pairs of vortices.

In the case of quantum fluids, vortex circulations are quantised and take only the values of $\pm h/m$, where $h$ is the Planck constant and $m$ is the mass of an atom.  When modelling with the GPE, the natural units to use are the \textit{healing length} units.  Distances are measured in terms of the healing length, $\xi=\hbar/\sqrt{m\mu}$, and times in terms of, $\tau=\hbar/\mu$, where $\mu$ is the chemical potential.  In these units, the circulations are $\pm 2\pi$.  We refer to a vortex of circulation $+2\pi$ as a vortex and a vortex of circulation $-2\pi$ as an antivortex. 

\subsection{A Dissipative PVM in a Doubly-Periodic Square Box}\label{subsec_diss_pvm} 
To model the dynamics of the vortices we use a doubly-periodic square box of length $L=2048\xi$.  To account for interactions between vortices across the periodic boundaries we use the following equations \cite{Weiss1991}:
\begin{equation}
\frac{d{\tilde{x}_{i}}}{dt}=\sum_{j=1,j\neq i}^{N_{v}} \kappa_j\sum_{m=-\infty}^{\infty}  \frac{-\sin(\tilde{y}_{ij})}{\cosh(\tilde{x}_{ij}-2\pi m)-\cos(\tilde{y}_{ij})}\, .
\label{eqn:pv_eqm_x_periodic}
\end{equation}
\begin{equation}
\frac{d{\tilde{y}_{i}}}{dt}=\sum_{j=1,j\neq i}^{N_{v}} \kappa_j\sum_{m=-\infty}^{\infty}  \frac{\sin(\tilde{x}_{ij})}{\cosh(\tilde{y}_{ij}-2\pi m)-\cos(\tilde{x}_{ij})}\, .
\label{eqn:pv_eqm_y_periodic}
\end{equation}
As above, the circulation of vortex $i$ is $\kappa_{i}$ but its coordinates, $(x_{i},y_{i})$, are rescaled by $2\pi/L$ to give $(\tilde{x}_{i},\tilde{y}_{i})$, so that in these units the box has a length of $2\pi$.  $\tilde{x}_{i}-\tilde{x}_{j}$ and $\tilde{y}_{i}-\tilde{y}_{j}$ are written as $\tilde{x}_{ij}$ and $\tilde{y}_{ij}$.  The time units are also scaled to give $\kappa_{i}=\pm 1$   The first sum is over all vortices except for vortex $i$ and the second is over the integer $m$.  Since the term in this second sum falls with increasing $m$, an excellent level of precision is achieved by curtailing the sum to be from $m=-10$ to $m=+10$. 

We use a Dormand-Prince 8th Order Runge-Kutta algorithm \cite{Prince1981} to solve these equations of motion, with an absolute tolerance of $1\times 10^{-8}$.  This gives us a non-dissipative velocity, $\mathbf{v}_{i}=(d\tilde{x}_{i}/dt,d\tilde{y}_{i}/dt)$, for each vortex.  To include the effects of dissipation we add a dissipative component at right angles to this to give the overall velocity, including dissipation, $\mathbf{v}_{i}^{\rm{diss}}=\mathbf{v}_{i}-\gamma\kappa_{i}\hat{\mathbf{z}}\times\mathbf{v}_{i}$, \cite{Billam2015}.  The values of $\gamma$ we use are $\{2^{-8},2^{-7},\,.\,.\,.\,,2^{-1},1,\infty \}$, where $\gamma=\infty$ is the purely dissipative case, in which $\mathbf{v}_{i}^{\rm{diss}}=-\kappa_{i}\hat{\mathbf{z}}\times\mathbf{v}_{i}$. 

A final, key aspect of vortex dynamics that we add to the simulation is the annihilation of vortices and antivortices that come into too close proximity.  This is observed in experiments and in GPE simulations and is the main mechanism by which phase ordering takes place.  To include this in our dissipative PVM, at each time-step we remove any vortex-antivortex pair that are within $1\xi$ of each other.

\subsection{Generating Initial Conditions}\label{subsec_gen_ics}
The initial arrangement of vortices and antivortices affects the dynamics that follow.  For example, an isolated vortex-antivortex pair, known as a dipole, will move in a straight line, perpendicular to the line joining the vortices.  Alternatively, a cluster made up only of vortices (or only of antivortices) will tend to rotate about its centre, so long as no other vortices are near.  The effect of dissipation on dipoles is to draw the constituent vortex and antivortex closer together over time, whereas it tends to scatter clusters apart.  Therefore, when generating multiple realizations of the initial conditions for repeat runs of the point vortex simulations it is vital to ensure they have simlar distributions of dipoles, clusters and free vortices.  To achieve this, we note that the Hamiltonian, $H_{\square}$, for a system of point vortices in a doubly-periodic square box \cite{Weiss1991}, gives an unambiguous parameter for measuring vortex configurations:
\begin{equation}
H_{\square}=-\sum_{i=1}^{N_{v}}\sum_{j=1,j \neq i}^{N_{v}} \frac{\kappa_i \kappa_j}{2}\sum_{m=-\infty}^{\infty} \ln{\left( \frac{\cosh{\tilde{x}_{ij}-2\pi m}-\cos{\tilde{y}_{ij}}}{\cosh{2\pi m}}\right)}-\frac{\tilde{x}_{ij}^2}{2\pi}\, .
\label{eqn:pv_H_periodic}
\end{equation}
In this equation all symbols have the same meanings as in Eq.~\ref{eqn:pv_eqm_x_periodic} \& \ref{eqn:pv_eqm_y_periodic}.  The double sum at the beginning adds contributions from all pairs of vortices and, as before, an excellent level of precision is achieved by curtailing the sum over $m$ to be from $m=-10$ to $m=+10$.  Calculating this value for the initial configuration and dividing the result by the number of vortices, $N_{v}$, gives the initial energy per vortex $E^{0}/N_{v}$.  A completely random configuration of vortices and antivortices would have $E^{0}/N_{v}=0$, increasing the amount of clustering of like-signed vortices increases $E^{0}/N_{v}$ and pairing vortices and antivortices  into dipoles makes $E^{0}/N_{v}$ more negative. 

To create initial conditions with specified values of $E^{0}/N_{v}$ we first scatter $2069$ vortices and $2069$ antivortices in the box.  Then we employ a biased random walk algorithm in which the positions of vortices are updated at random and these updated positions accepted only if they bring $E^{0}/N_{v}$ closer to the desired value.  Once the desired values is achieved to within a $1\%$ tolerance, further steps are undertaken in order to thermalize the configuration.  In each step, two vortices are randomly selected and moved by different, random amounts.  If $E^{0}/N_{v}$ remains within the tolerance the step is accepted.  Throughout all of this process, changes that bring vortices closer than $1\xi$ are rejected. 

Using this approach we create $100$ stochastic realizations of vortex coordinates for each value of $E^{0}/N_{v}=\{-4,-3,-2,-1,0.001,1,2,3,4\}\pm 1\%$.  These provide the initial conditions for our dissipative PVM.

\section{A Method for Determining the Scaling Region}\label{sec_analysis}
\subsection{Constructing the Velocity Field and Correlation Function}\label{subsec_vel_corr}
\begin{figure}[t]
\centering
\includegraphics[width=\textwidth]{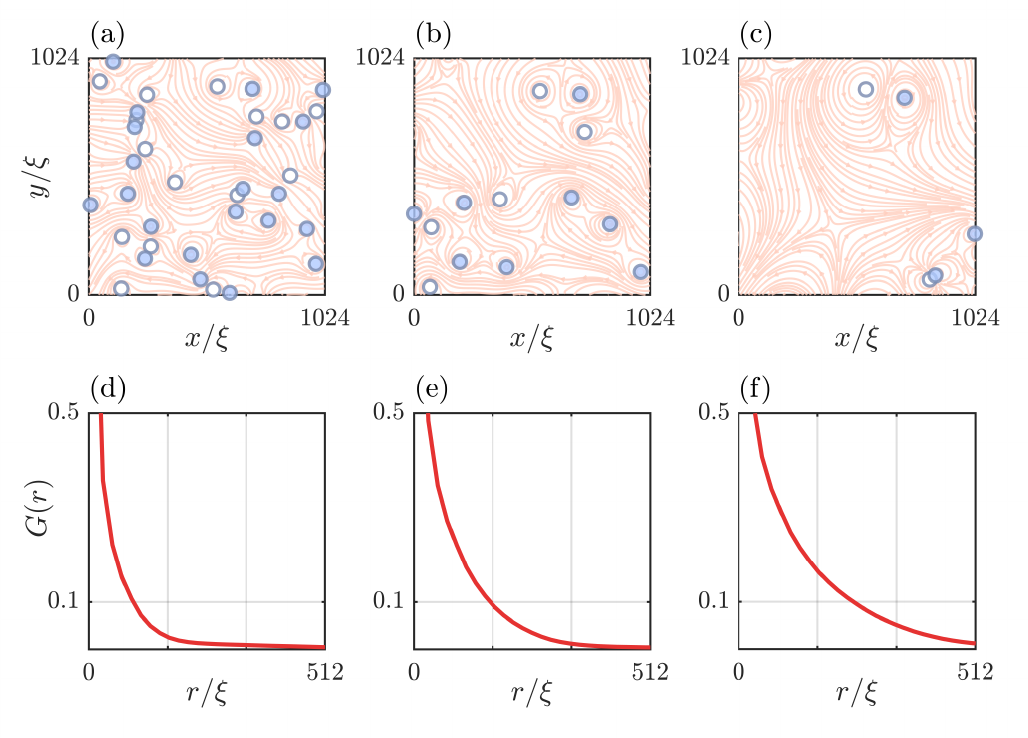}
\caption{Fluid velocity calculated from point vortex configurations and two-point correlation functions for the initial conditons $\gamma=1$, $E^{0}/N_{v}=4$.  (a)-(c)  Open (white) circles show the locations of vortices and filled (blue) circles show the locations of antivortices.  The pale red streamlines show the fluid velocity at the times $t=\{4200,13200,32700\}\tau$. (d)-(f) Are the two-point correlation function $G(r)$ calculated from the corrected speed at the same times.}\label{fig:velocity_field}
\end{figure}

In order to demonstrate dynamical scaling, a two-point correlation function is generally calculated using a relevant scalar order parameter.  In GPE simulations this may be the quantum mechanical phase or the fluid density, in the Ising model it is the magnetization, in our case we use the speed of the fluid. To find this we first construct the velocity field from the coordinates of the point vortices.  We consider a $100\times100$ grid of points $(\tilde{x}_{g},\tilde{y}_{g})$, filling the square box of length $2\pi$ and find the velocity of the fluid at each grid point, $\mathbf{v}=(v_{\tilde{x}},v_{\tilde{y}})$.  This is done by summing the contributions, at that location, of all of the point vortices, whilst accounting for the periodic boundary conditions \cite{Weiss1991}: 
\begin{equation}
v_{\tilde{x}}(\tilde{x}_{g},\tilde{y}_{g})=\sum_{i} \kappa_i\sum_{m=-\infty}^{\infty}  \frac{-\sin(\tilde{y}_{g}-\tilde{y}_{i})}{\cosh(\tilde{x}_{g}-\tilde{x}_{i}-2\pi m)-\cos(\tilde{y}_{g}-\tilde{y}_{i})}\,,
\label{eqn:pv_vel_x_periodic}
\end{equation}
\begin{equation}
v_{\tilde{y}}(\tilde{x}_{g},\tilde{y}_{g})=\sum_{i} \kappa_i\sum_{m=-\infty}^{\infty}  \frac{\sin(\tilde{x}_{g}-\tilde{x}_{i})}{\cosh(\tilde{y}_{g}-\tilde{y}_{i}-2\pi m)-\cos(\tilde{x}_{g}-\tilde{x}_{i})}\,.
\label{eqn:pv_vel_y_periodic}
\end{equation}
Eq.~\ref{eqn:pv_vel_x_periodic} \& \ref{eqn:pv_vel_y_periodic} are almost the same as the non-dissipative point vortex equations of motion, Eq.~\ref{eqn:pv_eqm_x_periodic} \& \ref{eqn:pv_eqm_y_periodic}, except for two differences; the sum is over all of the vortices and the coordinates, $(\tilde{x}_{g},\tilde{y}_{g})$, are those of a grid point and not those of a vortex.  This is because the point vortex equations arise from considering the contributions of all other vortices to the fluid velocity at that point and asserting that, in the absence of dissipation, the vortex moves at the local fluid velocity.  See Fig.~\ref{fig:velocity_field} (a)-(c) for examples of the velocity field constructed from the point vortex coordinates in this fashion.  

Using the velocity, $\mathbf{v}=(v_{\tilde{x}},v_{\tilde{y}})$, we find the corrected speed at each grid point, $v_{c}$, from the difference between the speed, $|\mathbf{v}|$, and the mean speed taken across the whole grid, $\langle|\mathbf{v}|\rangle$: 
\begin{equation}
v_{c}=|\mathbf{v}|-\langle |\mathbf{v}| \rangle\,.
\label{eqn:corr_speed}
\end{equation}
This is the scalar order parameter we use to find the two-point correlation function and ultimately provides the basis for demonstrating dynamical scaling.  

The correlation function, $G$, for the corrected speed, $v_{c}$, between two points separated by a displacement $(\tilde{x},\tilde{y})$, is given by: 
\begin{equation}
G(\tilde{x},\tilde{y})=\frac{\sum_{i}\sum_{j} v_{c}(\tilde{x}_{i},\tilde{y}_{j}) v_{c}(\tilde{x}_{i}-\tilde{x},\tilde{y}_{j}-\tilde{y})}{\sum_{i}\sum_{j} v_{c}(\tilde{x}_{i},\tilde{y}_{j})^{2}}\,. 
\label{eqn:corrn_func_norm}
\end{equation}
Where the sum over $i$ is across all grid points in the $\tilde{x}$ direction, the sum over $j$ is for all those in the $\tilde{y}$ direction and the denominator ensures normalization such that $G(0,0)=1$.  The Wiener-Khinchin theorem \cite{Champeney1987} provides a more succint, and computationally more efficient, formula in terms of Fourier transforms:
\begin{equation*}
G(\tilde{x},\tilde{y})=\mathcal{F}^{-1}[|\mathcal{F}(v_{c})|^2]\,. 
\label{eqn:corrn_func_fourier}
\end{equation*}
To convert from $G(\tilde{x},\tilde{y})$ to a one-dimensional $G(\tilde{r})$ we consider nested annuli of thickness, $\delta \tilde{r}$, where $\delta \tilde{r}$ is the distance between adjacent grid points and $\tilde{r}^{2}=\tilde{x}^{2}+\tilde{y}^{2}$.  We then find $G(\tilde{r})$ by calculating the mean value of $G(\tilde{x},\tilde{y})$ across all the grid points in the annulus from $\tilde{r}$ to $\tilde{r}+\delta \tilde{r}$.  Finally, we multiply the distances by $L/2\pi$ to return to the original length scale used for our system and recover $G(r)$.  We repeat this for each time-step, for all $100$ stochastic realizations and for the whole ensemble of initial conditions.

\subsection{Rescaling the Correlation Function and Testing for Dynamical Scaling}\label{subsec_recale_overlap}
The theory of dynamical scaling predicts that the correlation function, $G(r,t)$ should vary only by a change in length scale as time progresses.  If this is the case, finding a suitable length at each time-step and rescaling the distances by this length, should cause the plots of $G(r)$ at different times to collapse onto a single curve.

The process we use to rescale the correlation function for a given set of initial conditions is as follows:
\begin{enumerate}
	\item Consider one time-step and find the mean values of $G(r)$ across the $100$ stochastic realizations (from here on $G$ should be understood as the average over these realizations). 
	\item Use cubic spline interpolation to find the radius at which $G(r)=0.1$.  We define this as the correlation length at this time-step, $L_{c}(t)$. (We use $0.1$ for our data as it gives sufficient points, at both smaller and larger radii, to ensure a reliable interpolation).
	\item Rescale the values of radius by dividing them by $L_{c}(t)$.
	\item Repeat the steps 1-3 for all time-steps.	
\end{enumerate}

\begin{figure}[t]
	\centering
	\includegraphics[width=\textwidth]{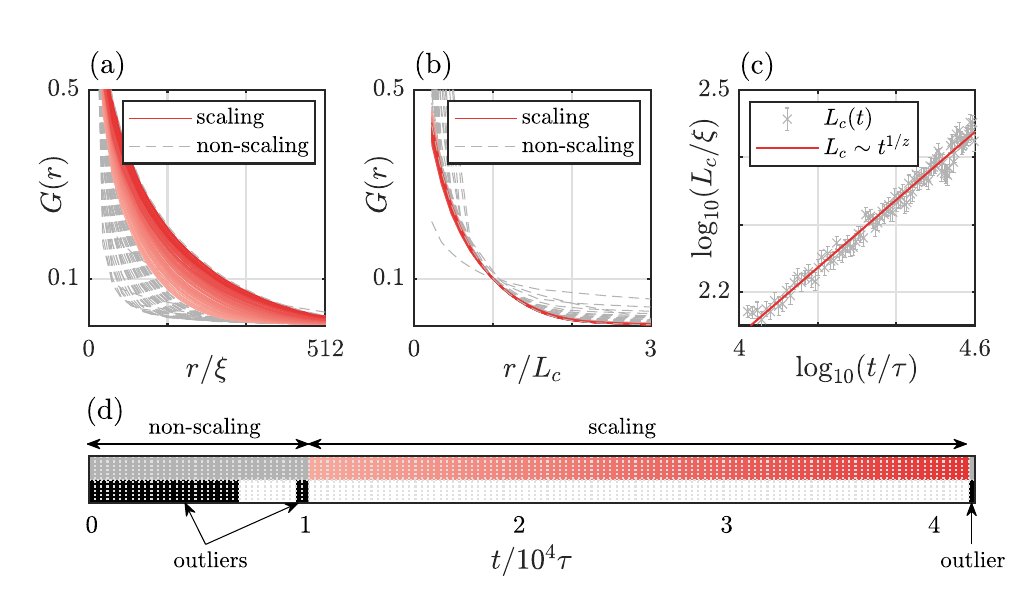}
	\caption{Identifying the scaling region and determing $z$ for the initial conditions $\gamma=1,\,E^{0}/N_{v}=4$.  (a) Shows the two-point correlation function $G(r)$ for each time-step.  (b) Shows  $G(r)$ with the distances rescaled by $L_{c}$, where $L_{c}$ is the value of $r$ when the unscaled $G(r)=0.1$.  In both (a) and (b) those time-steps identified as within the scaling region are plotted as solid red lines, getting darker as time progresses, and those outside of the scaling region are plotted as gray dashed lines.  In the rescaled plots, (b), all curves in the scaling region overlap.  (c) Shows the data for correlation length against time plotted as grey crosses, the bars show the standard error in $L_{c}$ calculated across the $100$ stochastic realizations at each time-step.  The red line is a power law $L_{c}\sim t^{1/z}$, with $z=1.985\pm0.003$ determined using a bootstrap method \cite{Efron1993} in which we resample $L_{c}(t)$ $100$ times, conduct a power law fit each time and find the mean and standard error in $z$ across the resamplings.  (d)  Illustrates the time-steps identified as outliers as black blocks in the bottom line and those in the scaling region as red blocks on the top line.}\label{fig:G_r_outliers}
\end{figure}

This gives arrays of data for $G$ and $r/L_{c}$ at each time-step.  Simply plotting $G$ against $r/L_{c}$ allows us to check visually if the curves for different time-steps collapse.  However, a more systematic approach is desired to determine the start and end of any period of dynamical scaling in a robust and reliable fashion.  The method we implement to test for collapse of the different curves is:
\begin{enumerate}
	\item For every time-step, use spline interpolation to find the values of $G$ on the same uniform grid of values of $r/L_{c}$.  Use $N_{g}$ points in the range from the smallest value of $r/L_{c}(t)$ for any time-step up to $r/L_{c}(t)=5$.  By this point the values of $G$ should be very close to zero in all cases, so continuing further is unnecessary.  We use $N_{g}=40$.
	\item For each time-step there are now $N_{g}$ values of $G$.  Therefore each time-step can now be represented by a point in an $N_{g}$-dimensional Euclidean space:
	\begin{equation}
	\mathbf{G}^{i}=(G_{1}^{i},G_{2}^{i},...,G_{N_{g}}^{i})\,,
	\end{equation}
	 where, $\mathbf{G}^{i}$ is the vector position representing the correlation function at timestep $i$ and $(G_{1}^{i},G_{2}^{i},...,G_{N_{g}}^{i})$ are the coordinates in each of the dimensions (i.e. the values of $G$ on the grid of $N_{g}$ radii).  The problem is now one of determining which time-steps are outliers and which are not, see Fig.~\ref{fig:MD_explanation}.  
	 
	 \begin{figure}[t]
	 	\centering
	 	\includegraphics[width=\textwidth]{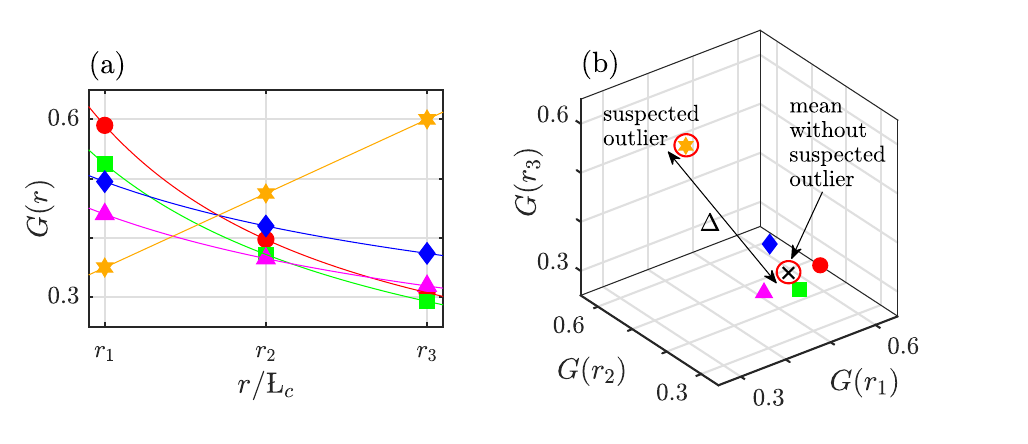}
	 	\caption{Representing curves as points in an $N_{g}$-dimensional Euclidean space.  (a) Five curves $G(r)$ plotted on a grid of three points ${r_{1},r_{2},r_{3}}$.  (b) The same curves represented by points in a three-dimensional Euclidean space.  Similar curves, such as those shown with red circles, green squares, blue diamonds and magenta triangles, become points that are clustered close together; whilst a very different curve, like the one shown with orange stars, becomea a point very far from the others.  To identify an outlier, we need to quantify the deviation, $\Delta$, of a suspected point from the mean position of all the other points.  We use the Mahalanobis distance, in the $N_{g}$-dimensional Euclidean space, to do this. }\label{fig:MD_explanation}
	 \end{figure}
 
	\item Find the mean, $\boldsymbol{\mu}$, and the covariance, $\mathbf{C}$, of $\mathbf{G}^{i}$ across all of the time-steps.  $\boldsymbol{\mu}$ is a vector of length $N_{g}$ and $\mathbf{C}$ is a matrix of size $N_{g}\times N_{g}$.
	\item The squared deviation of a point, $\mathbf{G}^{i}$, from the mean, $\boldsymbol{\mu}$,  normalized by the covariance between the data from the orginal curves, is given by the Mahalanobis distance squared, $MD_{i}^{2}$, \cite{Mahalanobis1936_2018}:
	\begin{equation}
	MD_{i}^{2}=(\mathbf{G}^{i}-\boldsymbol{\mu}) \mathbf{C}^{-1}(\mathbf{G}^{i}-\boldsymbol{\mu})^{T}\,.
	\label{eqn:Mahalanobis_dist}
	\end{equation}  
	Calculate $MD_{i}^{2}$ for each time-step and order them from largest to smallest.  Then, to identify and eliminate outliers:
	\item Choose the time-step with the largest value of $MD_{i}^{2}$ and recalculate $\boldsymbol{\mu}$ and $\mathbf{C}$ without it (i.e. treate it as an outlier).
	\item Recalculate $MD_{i}^{2}$ for the suspected outlier using the new $\boldsymbol{\mu}$ and $\mathbf{C}$.  If the value is below a cutoff (see discussion below), this timestep is not an outlier and the process is stopped.  If it is above the cutoff, this timestep is identified as an outlier and removed from future consideration.
	\item Recalculate $MD_{i}^{2}$ for all the other points, with the previously identified outlier(s) excluded and order these values from largest to smallest.
	\item Identify the time-step with the new largest value of $MD_{i}^{2}$ and repeat steps 5-7. 
	\item Continue iteratively until a suspected outlier has $MD_{i}^{2}$ less than the cutoff, at which point all outliers have been identified.
	\item All remaining timesteps are considered to overlap.  The scaling region is then the longest stretch of uninterrupted timesteps not containing any outliers.  See Fig.~\ref{fig:G_r_outliers} for an example.
\end{enumerate}

The choice of cutoff for determining outliers is a slighlty subtle one.  If the values of $\mathbf{G}^{i}$ at each radius are independent and normally distributed then $MD_{i}^{2}$ would qualify as a $\chi^{2}$ statistic and the critical $\chi^{2}$ value for a suitable significance level would be an appropriate choice.  However, our data are not normally distributed, so instead we use a cutoff of $4\times N_{g}$, where $N_{g}$ is the number of grid-points.  The logic behind this decision is that for the case in which $\mathbf{G}^{i}$ at each radius varies independently of every other radius, the covariance matrix is diagonal and $MD_{i}^{2}$ becomes a sum of similar terms:
\begin{equation}
MD_{i}^{2}=(\mathbf{G}^{i}-\boldsymbol{\mu}) 
\begin{pmatrix}
1/\sigma_{1}^{2} & 0 &  . & . & 0\\
0 & 1/\sigma_{2}^{2} & . & . & 0\\
. & . & . & . & .\\
. & . & . & . & .\\
0 & 0 & . & . & 1/\sigma_{N_{g}}^{2}
\end{pmatrix}
(\mathbf{G}^{i}-\boldsymbol{\mu})^{T}\,= \sum_{k=1}^{N_{g}}\left(\frac{G^{i}_{k}-\mu_{k}}{\sigma_{k}}\right)^2\,.
\label{eqn:Mahalanobis_dist_simple}
\end{equation} 
So, if a hypothetical curve has every point one standard deviation from the mean, then $MD^{2}=\sum_{k=1}^{N_{g}}1^{2}=N_{g}$, whereas if each point is two standard deviations from the mean then $MD^{2}=\sum_{k=1}^{N_{g}}2^{2}=4N_{g}$.  We consider this a reasonable choice of cutoff, whilst recognizing that our values of $\mathbf{G}^{i}$ at each radius are not, in general, independent.  However, a lower value could be used if a stricter definition of collapse is preferred.  The cutoff is the one free parameter in this approach and plays a similar role to the significance level in hypothesis testing.  Fig.~\ref{fig:z_values} (a) shows how the choice of cutoff affects the percentage of timesteps identified as within the scaling region and consequently the values of dynamical critical exponent determined from the scaling region data.  

\subsection{The Dynamical Critical Exponent}\label{subsec_z}
\begin{figure}[t]
	\centering
	\includegraphics[width=\textwidth]{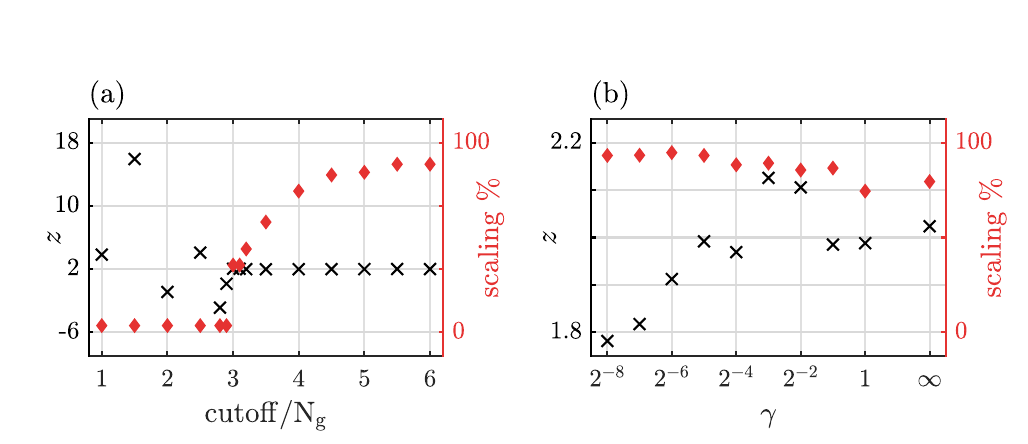}
	\caption{Determining the dynamical critical exponent, $z$, for $E^{0}/N_{v}=4$.  (a)  The black crosses (left-hand axis) show how the value of $z$ varies with the choice of cutoff for $\gamma=1$  (errorbars are not shown as they are either too large, for cutoff $< 3N_{g}$, or too small, for cutoff $\geq 3N_{g}$).  The red diamonds (right-hand axis) show how the \% of timesteps identified as within the scaling region varies.  Using a cutoff of less than $3N_{g}$ is problematic as only a tiny fraction of timesteps are within the scaling region and the fit is unreliable.  (b)  The black crosses (left-hand axis) show how the value of $z$ varies with the dissipation, $\gamma$, for a cutoff of $4N_{g}$ (errorbars would be too small to see on this scale).  The red diamonds (right-hand axis) show the corresponding \% of timesteps in the scaling region.  A large scaling region is identified for all $\gamma$ and $z\approx2$ but dependent upon $\gamma$, a qualitatively similar result to that reported in \cite{Groszek2021}.  In all cases we determine $z$ using the same bootstrap method described in Fig.~\ref{fig:G_r_outliers}.  }\label{fig:z_values}
\end{figure}

To calculate values for the dynamical critical exponent, $z$, we conduct a least squares fit of the power law $L_{c}\sim t^{1/z}$ to the data  $L_{c}(t)$ for just the times within the scaling region.  By identifying the scaling region in a robust fashion, with a suitable choice of cutoff, prior to conducting the fit, we can be more confident that the predicted power law relationship is valid and the value of $z$ found is an accurate measure of the dynamical critical exponent of the system.  Fig.~\ref{fig:z_values} (b) shows how the value of $z$, determined in this way, varies with the level of dissipation, $\gamma$.   

\section{Conclusions}\label{sec_conc}
We have described a systematic method, based on the concept of Mahalanobis distance, for identifying outliers when comparing graphs of rescaled correlation functions.  By eliminating those times for which the curves do not collapse onto one single curve, our approach gives a reliable identification of the scaling region.  We have also demonstrated its application to the phase-ordering behaviour of a quantum fluid following a rapid quench, using data generated using a dissipative PVM.  This produced results that correspond qualitatively and quantitatively with those from other studies that make use of GPE simulations and visual identification of the scaling region (with subsequent checks) \cite{Groszek2021,Karl2017}.  In particular, we found values of dynamical critical constant that are close to two but depend on dissipation.  By applying this method we can be confident that the data used to determine the dynamical critical exponent came only from times during which the system displays scaling behaviour.  Further work could include an application of this approach to the correlation functions for a wider ensemble of initial conditions, to the results of GPE simulations or to studies of non-equilibrium behaviours of systems outside of the field of quantum fluids.

\backmatter

%
%
%
%
%
%
%
\section*{Declarations}
%
%
\begin{itemize}
\item \textbf{Funding} This work was supported by a Lady Bertha Jeffreys PhD Studentship.
\item \textbf{Competing interests} The authors have no competing interests to declare that are relevant to the content of this article.
\item For the purpose of open access, the author has applied a Creative Commons Attribution (CC BY) licence to any Author Accepted Manuscript version arising from this submission.
\end{itemize}

\bibliography{RT_QFS_bibliography}

\end{document}